\documentclass[conference]{IEEEtran}
\IEEEoverridecommandlockouts
\usepackage{cite}
\usepackage{amsmath,amssymb,amsfonts}
\usepackage{algorithmic}
\usepackage{graphicx}
\usepackage{textcomp}
\usepackage{algorithm}
\usepackage{bm}
\usepackage{amsthm}
\usepackage{amsmath}
\newtheorem{remark}{Remark}
\usepackage{url}
\usepackage{hyperref}
\usepackage{xcolor}
\def\BibTeX{{\rm B\kern-.05em{\sc i\kern-.025em b}\kern-.08em
    T\kern-.1667em\lower.7ex\hbox{E}\kern-.125emX}}
\begin{document}

\title{\LARGE\textbf{~\\~\\Trajectory planning for multiple autonomous underwater vehicles with safety guarantees}\\
\thanks{S. Zhang, Y. Yang, S. Siriya and Y. Pu are with the Department of Electrical and Electronic Engineering, University of Melbourne. \{shuhaoz2@student., yujyang1@student., ssiriya@student., ye.pu@\}unimelb.edu.au}
}
\author{\IEEEauthorblockN{Shuhao Zhang, Yujia Yang, Seth Siriya, Ye Pu}}

\maketitle

\begin{abstract}
This paper addresses the trajectory planning for multiple autonomous underwater vehicles (AUVs) in strong waves that can disturb the AUVs' trajectory tracking ability and cause obstacle and inter-vehicle collisions. A novel approach based on Hamilton-Jacobi differential game formulation and model predictive control is proposed to generate safety-guaranteed trajectories for AUV systems considering wave disturbance that is either time-varying state-independent or time-varying state-dependent. This approach plans safety-guaranteed trajectories for multiple AUVs and considers the shapes of AUVs, where nonlinear smooth collision avoidance constraints are obtained via strong duality.
\end{abstract}
% \begin{IEEEkeywords}
% Underwater vehicle, path planning, trajectory planning, 
% \end{IEEEkeywords}

\section{Introduction}
In ocean-based industries, AUVs have great potential in missions such as marine exploration, ocean search and rescue, and coastal surveillance, whether it be individually or as a fleet of AUVs cooperatively. However, ocean disturbances like strong waves and currents pose challenges to the navigation of AUVs. The existing planning and control methods are not sufficient to provide real-time safety-guaranteed trajectory planning for a single AUV or multiple AUVs.

The majority of earlier works on trajectory planning for multiple AUVs, such as \cite{bb1}, which presented a two-stage cooperative trajectory planner for obstacle and inter-vehicle collision avoidance, did not consider strong disturbances and dynamic obstacles.
Dynamic programming is commonly used in multi-AUV trajectory planning problems that consider time-varying wave disturbances \cite{bb2}, \cite{bb3}, but is not fast enough for online planning.
Model predictive control (MPC) can use wave models to account for wave disturbances, which showed good performance for single AUV control but with no safety guarantees \cite{bb4}. The work in \cite{bb5} achieved motion control and collision avoidance of multiple AUVs using MPC, but did not consider strong disturbances and dynamic obstacles.

Recently, a real-time trajectory-planning modular framework, FaSTrack, was proposed to provide safety-guaranteed trajectory planning with bounded disturbances \cite{b2} through Hamilton-Jacobi (HJ) differential game formulation. Building on this, \cite{b1} combined FaSTrack with MPC to achieve safety-guaranteed single AUV trajectory planning considering various wave disturbance models. Compared to the tube-based MPC planning method in \cite{bb6} which considers nonlinear system uncertainties and external disturbances, the HJ method combined with an MPC formulation shows potential in safety-guaranteed real-time trajectory planning for multiple AUVs considering more complex disturbance models. Inspired by the collision avoidance reformulation in \cite{b3}, this work considers the shapes of AUVs, instead of the point-mass AUV models used in existing AUV collision avoidance methods.

Building on the single-vehicle real-time safety-guaranteed trajectory planning method presented in \cite{b1}, this work makes the following main contributions
\begin{itemize}
    \item We propose a generalized approach based on the HJ method to do safety-guaranteed trajectory planning for multiple AUVs considering various wave disturbance models, such as time-varying and state-dependent ones.
    \item We show the ability of our approach to handling dynamic obstacles by presenting reformulations of the collision avoidance constraints via the Lagrange duality approach to consider the shapes of AUVs.
    \item We demonstrate our approach in two trajectory planning scenarios: a single-AUV one, and a multi-AUV one considering various wave disturbance models.
\end{itemize}
%\clearpage

\section{Preliminaries}
\subsection{Notation}
% Let $\mathcal{K},\mathcal{K}^*\in\mathbb{R}^l$ denote a proper cone and its dual cone. For two vectors $a,b\in\mathbb{R}^n$, $(b-a)\in\mathcal{K}$ is equivalent to $a\preceq_Kb$. Let $\|\cdot\|_*$ denote the dual norm of $\|\cdot\|$. 
Let $\mathcal{K}\in\mathbb{R}^l$ denote a proper cone. For two vectors $a,b\in\mathbb{R}^l$, $(b-a)\in\mathcal{K}$ is equivalent to $a\preceq_\mathcal{K}b$.
The Euclidean norm is denoted by $\|\cdot\|_2$. The quadratic norm with respect to a positive definite matrix $Q$ is defined by $\|x\|^2_Q=x^\top Qx$. Let $\oplus$ and $\ominus$ denote the Minkowski sum and difference, respectively. 
%The Minkowski sum of two sets is defined by $\mathbb{W}\oplus\mathbb{Z}:=\{w+z:w\in\mathbb{W},z\in\mathbb{Z}\}$.
\subsection{Models for AUVs and wave disturbance}\label{S_II_B}
We introduce a 2D AUV model (\ref{AUV_model}) moving in the $x$-$z$ plane under wave disturbance, which is proposed in \cite{b1} and is simplified from the 6 degree-of-freedom (DOF) model presented in \cite{b6} and \cite{b7}. Let $s=[x,z,u_r,w_r]^\top$ denote the states of the AUV dynamics, where $x,z$ denote the position and $u_r,w_r$ denote the velocity in the inertial $x$ and $z$ coordinates, respectively. Let $u_s = [T_A, T_B]^\top$ denote the control forces of the AUV dynamics. The model uncertainties are considered as unknown additive bounded noises $d_x$, $d_z$, $d_{u_r}$, and $d_{w_r}$. The system dynamics is described as
\begin{equation} \allowdisplaybreaks
    \begin{aligned} \allowdisplaybreaks
        \Dot{x} &= u_r + V_{f,x} + d_x, \\ 
        \Dot{z} &= w_r + V_{f,z} + d_z, \\
        \Dot{u}_r &= (m-X_{\Dot{u}_r})^{-1}((\Bar{m} - m)A_{f,x}, \\
        &\quad\quad-(X_{u_r}+X_{|u_r|u_r}|u_r|)u_r + T_A) + d_{u_r}, \\
        \Dot{w}_r &= (m-Z_{\Dot{w}_r})^{-1}((\Bar{m} - m)A_{f,z}-(-g(m - \Bar{m})), \\
        &\quad\quad-(Z_{w_r}+Z_{|w_r|w_r}|w_r|)w_r + T_B) + d_{w_r}. \allowdisplaybreaks
    \end{aligned}
    \label{AUV_model}
\end{equation}
The systems parameters in (\ref{AUV_model}) are the vehicle mass $m$, the displaced fluid mass $\Bar{m}$, the additive masses $X_{\Dot{u}_r}$, $Z_{\Dot{w}_r}$, the linear damping factors $X_{u_r}$, $Z_{w_r}$, and the quadratic damping factors $X_{|u_r|u_r}$, $Z_{|w_r|w_r}$. Moreover, $g$ denotes the gravitational acceleration. The velocity components of the wave disturbance $[V_{f,x},V_{f,z}]^\top$ and their derivative with respect to time $[A_{f,x},A_{f,z}]^\top$ shown in (\ref{AUV_model}) reflect the fact that waves can simultaneously affect both the position and the velocity of the AUV. Two types of wave disturbance models are proposed in \cite{b1}:
\subsubsection{TVSD wave disturbance model}
One is the time-varying state-dependent (TVSD) wave disturbance model
\begin{equation}
    \begin{bmatrix}
    V_{f,x}(x,z,t)\\
    V_{f,z}(x,z,t)
    \end{bmatrix}=A\omega e^{-kz}
    \begin{bmatrix}
    \cos(kx-\omega t)\\
    -\sin(kx-\omega t)
    \end{bmatrix},
    \label{TVSD_V}
\end{equation}
and its derivative with respect to time
\begin{equation}
    \begin{bmatrix}
    A_{f,x}(x,z,t)\\
    A_{f,z}(x,z,t)
    \end{bmatrix}=A\omega^2e^{-kz}
    \begin{bmatrix}
    \sin(kx-\omega t)\\
    \cos(kx-\omega t)
    \end{bmatrix},
    \label{TVSD_A}
\end{equation}
where $A$ denotes the wave amplitude, $\omega$ denotes the wave frequency, and $k=\omega^2/g$ denotes the wave number.
\subsubsection{TVSI wave disturbance model} When the effect of the disturbance on the vehicle is similar over the entire bounded region of positions, the time-varying state-independent (TVSI) wave disturbance model
\begin{equation}
    \begin{bmatrix}
    V_{f,x}(t)\\
    V_{f,z}(t)
    \end{bmatrix}=\Tilde{A}_v
    \begin{bmatrix}
    \cos(\Tilde{\phi}_v-\omega t)\\
    -\sin(\Tilde{\phi}_v-\omega t)
    \end{bmatrix} + 
    \begin{bmatrix}
    d_{f,v,x}\\
    d_{f,v,z}
    \end{bmatrix},
    \label{TVSI_V}
\end{equation}
and its derivative with respect to time
\begin{equation}
    \begin{bmatrix}
    A_{f,x}(t)\\
    A_{f,z}(t)
    \end{bmatrix}=\Tilde{A}_a
    \begin{bmatrix}
    \sin(\Tilde{\phi}_a-\omega t)\\
    \cos(\Tilde{\phi}_a-\omega t)
    \end{bmatrix} + 
    \begin{bmatrix}
    d_{f,a,x}\\
    d_{f,a,z}
    \end{bmatrix},
    \label{TVSI_A}
\end{equation}
are considered approximation of (\ref{TVSD_V}) and (\ref{TVSD_A}), respectively. The parameters $\Tilde{A}_v$, $\Tilde{A}_a$, $\Tilde{\phi}_v$ and $\Tilde{\phi}_a$ depend on the wave amplitude and the wavenumber, and 
$d_{f,v,x}$, $d_{f,v,z}$, $d_{f,a,x}$ and $d_{f,a,z}$ denote the bounded uncertainty due to loss of spatial information in the wave disturbance model. We refer to \cite{b1} for a detailed discussion about the bounded relationship between the two wave disturbance models.
\begin{remark}
The pitch angle (rotation around the $y$-axis) of the AUV is ignored in the above model by assuming it is small, slow-changing, and self-regulating for the AUV application we study. 
\end{remark}

\subsection{HJ Reachability with Time-varying Dynamics}\label{S_II_C}
This subsection briefly introduces the definition of the HJ differential game presented in FaSTrack \cite{b2} and its earlier work \cite{b4}, which was also used for the safety-guaranteed single AUV trajectory planning in \cite{b1}. To formulate this differential game, we first define three systems used throughout the paper: the tracking, planning, and relative systems.

The \textit{tracking system} $\Dot{s}=f(s(t),u_s(t),d(t),t)$ represents the dynamics of the vehicle, where $s\in\mathbb{R}^{n_s}$, $u_s\in\mathcal{U}_s$ and $d\in\mathcal{D}$ denote the tracking states, the tracking control inputs and the tracking system disturbances, respectively. In this paper, the AUV model (\ref{AUV_model}) is used for the tracking system.

The \textit{planning system} $\Dot{p} = h(p(t),u_p(t))$ is a low complexity model used for the online trajectory planning. Consider a simple 2D AUV model
\begin{equation}
    \Dot{p}=h(p(t),u_p(t))=
    [
    \Dot{x}_p,\Dot{z}_p
    ]^\top=
    [
    u_{p,x},u_{p,z}
    ]^\top,\\
    \label{planning_system}
\end{equation}
where $p=[x_p,z_p]^\top\in\mathbb{R}^{n_p}$ is the planning states containing the planning system's positions in the inertial $x$-$z$ coordinate, and $u_p=[u_{p,x},u_{p,z}]^\top\in\mathcal{U}_p$ contains the control inputs of the planning system.

The \textit{relative system} representing the state difference between the tracking system and the planning system is defined as follows
\begin{equation}
    \Dot{r} = g(r(t),u_s(t),u_p(t),d(t),t),
    \label{eq_relative}
\end{equation}
where $g(\cdot,\cdot,\cdot,\cdot,\cdot)$ is a function of the relative states $r(t)$, the tracking inputs $u_s(t)$, the planning inputs $u_p(t)$, and the disturbance $d(t)$.
The relative states $r=[e^\top,\eta^\top]$ include the error states $e$, representing the tracking error between the tracking and the planning system, and the auxiliary states $\eta$, representing the remaining states that do not contribute to $e$. Furthermore, the relative states have slightly different definitions for the system with TVSI wave disturbance model and the system with TVSD wave disturbance model when using the simple AUV model in \eqref{planning_system} as the planning system. For the 2D AUV model (\ref{AUV_model}) with TVSI wave disturbance model (\ref{TVSI_V}) and (\ref{TVSI_A}), the relative states can be defined by
\begin{equation}
    r_1 = s - Mp,
    \label{relative_SI}
\end{equation}
where matrix $M$ augments the planning states to be consistent with the dimension of the tracking states as follows:
\begin{equation}
    M=\begin{bmatrix}
    1&0&0&0\\
    0&1&0&0
    \end{bmatrix}^\top.
    \label{M_matrix}
\end{equation}
However, for the 2D AUV model (\ref{AUV_model}) with TVSD wave disturbance model (\ref{TVSD_V}) and (\ref{TVSD_A}), two more states $
[x,z]^\top$, which are the true position of the tracking system, should be included, resulting in the following relative states.
\begin{equation}
    r_2 = \begin{bmatrix}
        s - Mp\\
        x\\
        z
    \end{bmatrix}.
    \label{relative_SD}
\end{equation}
We refer to \cite{b1} for the concrete formulations of the relative system with TVSI wave disturbance $\Dot{r}_1=g_1(r_1(t),u_s(t),u_p(t),d(t),t)$ derived from (\ref{relative_SI}), and the relative system with TVSD wave disturbance $\Dot{r}_2=g_2(r_2(t),u_s(t),u_p(t),d(t),t)$ derived from (\ref{relative_SD}).

The FaSTrack framework guarantees that the tracking system will always remain within an invariant set around the planning system.
The invariant set named the \textit{tracking error bound} (TEB) is derived from the value function of the differential game with respect to the relative system (\ref{eq_relative}), whose value function is defined as follows. 
\begin{equation}
\begin{aligned}
    &V(r,t)=\sup_{\gamma_p\in\Gamma_p(t),\gamma_d\in\Gamma_d(t)}\inf_{u_s(\cdot)\in\mathbb{U}_s(t)}\{\\
    &\quad\max_{\tau\in[t,T]}l(\xi_g(\tau;r,t,u_s(\cdot),\gamma_p[u_s](\cdot),\gamma_d[u_s](\cdot)))\},
\end{aligned}
\end{equation}
where $\gamma_p:\mathcal{U}_s\to\mathcal{U}_p$ and $\gamma_d:\mathcal{U}_s\to\mathcal{D}$ map tracking inputs to planning inputs and disturbance, respectively. The sets $\Gamma_p(t)$ and $\Gamma_d(t)$ are non-anticipative mapping sets, and $\mathbb{U}_s$ is the set of functions satisfying tracking inputs constraints for all time. The cost function 
$\max_{\tau\in[t,T]}l(\xi_g(\tau;r,t,u_s(\cdot),\gamma_p[u_s](\cdot),\gamma_d[u_s](\cdot)))$
denotes the maximum relative tracking error over time between the tracking and the planning system. Therein, $l(r)$ is set to be the Euclidean norm applied to $e$, and $r = \xi_g(\tau;r,t,u_s(\cdot),\gamma_p[u_s](\cdot),\gamma_d[u_s](\cdot))$ is the trajectory of (\ref{eq_relative}), which starts at state $r$ at time $t$ under tracking control $u_s(\cdot)$, planning control $\gamma_p[u_s](\cdot)$, and disturbance $\gamma_d[u_s](\cdot)$. (see \cite{b1} for more details).

The optimal tracking input $u_s^*(r,t)$ can guarantee every level set of the value function $V(r,t)$ is an reachable set. 
The reachability property of the value function of the differential game is used to derive the TEB.
Let the level set be $\mathcal{B}(t,V') = \{r:V(r,t)\leq V'\}$.
The minimum level $\underline{V}$ that involves the internal states $\eta_0$ at $t=0$ is
\begin{equation}
    \underline{V}:=\min_{V'}V'\quad \mathrm{s.t.}\quad\exists e\in\mathbb{R}^2,\begin{bmatrix}
    e^\top,\eta_0^\top
    \end{bmatrix}^\top\in\mathcal{B}(0,V').
    \label{min_V}
\end{equation}
The invariant set in the space of the error states, or the TEB, is as follows
\begin{equation}
    \mathcal{B}_{e,1}(t)=\{e:\exists\eta, V([e^\top,\eta^\top]^\top,t)\leq\underline{V}\},
    \label{TEB_SI}
\end{equation}
which provides a safety-guarantee for the relative system with the TVSI wave disturbance derived from (\ref{relative_SI}). The TEB for the relative system with TVSD wave disturbance derived from (\ref{relative_SD}) is
\begin{equation}
    \mathcal{B}_{e,2}(x,z,t)=\{e:\exists\eta, V([e^\top,\eta^\top,x,z]^\top,t)\leq\underline{V}\}.
    \label{TEB_SD}
\end{equation}
The value function is computed by solving the HJ partial differential equation
\begin{equation}
    \begin{aligned}
        \max&\Big\{\frac{\partial V}{\partial t} + \min_{u_s\in\mathcal{U}_S}\max_{u_p\in\mathcal{U}_p,d\in\mathcal{D}}\Big[\frac{\partial V}{\partial r}\cdot g(r,u_s,u_p,d,t)\Big],\\
        &l(r)-V(r,t)\Big\}=0,\, t\in[0,T],\, V(r,T)=l(r).
        \label{Calculate_TEB}
    \end{aligned}
\end{equation}
We refer to \cite{b1},\cite{b2} and \cite{b4} for more details of the value function and proof of the invariance property.

\section{MPC-Based Multi-Vehicle Trajectory Planning}
The multi-vehicle trajectory planning is formulated as an MPC optimization problem, which can later be applied to multi-AUV systems with wave disturbances.
\begin{subequations}
\begin{align}
    \min_{\bm{p^i},\bm{u_p^i} \: \forall i \in \mathcal{M}}&
    % \sum\limits_{i=1}^{M}\bigg(\sum\limits_{k=0}^{N-1}\Big(\|p_{k,i}-p_{r,i}\|^2_Q+\|u_{k,i}\|^2_R\Big)\notag\\
    % &\quad\quad\quad\quad\quad\quad\quad\quad\quad\quad+\|p_{N,i}-p_{r,i}\|^2_{Q_N}\bigg)\label{MPC_multia}\\
    \sum\limits_{i=1}^{M}\bigg(\sum\limits_{k=0}^{N}\|p_k^i-p_r^i\|^2_Q+ \sum\limits_{k=0}^{N-1}\|u_{p,k}^i\|^2_R\bigg)\label{MPC_multia}\\
    s.t.\quad&p_0^i\in\mathcal{I}_p^i,p_N^i\in\mathcal{G}_p^i,p_k^i\in\mathcal{P}_{p,k}^i,\label{MPC_multib1}\\
    &u_{p,k}^i\in\mathcal{U}_p^i,\label{MPC_multic}\\
    &p_{k+1}^i = h_d(p_{k}^i,u_{p,k}^i),\label{MPC_multid}\\
    % &\mathbb{C}_{k,i}^a(p_{k,i},t_k)\cap\mathbb{O}_q(t_k)=\emptyset,\label{MPC_multie}\\
    % &\mathbb{C}_{k,i}^a(p_{k,i},t_k)\cap\mathbb{C}_{k,j}^a(p_{k,j},t_k)=\emptyset, \label{MPC_multif}\\
    &\mathbb{C}(p_k^i)\cap\mathbb{O}^q=\emptyset,\forall i\in\mathcal{M}, q\in\mathcal{Q},\label{MPC_multie}\\
    &\mathbb{C}(p_k^i)\cap\mathbb{C}(p_k^j)=\emptyset,\forall i,j\in\mathcal{M}, i\neq j \label{MPC_multif}
\end{align}
    \label{MPC_multi}
\end{subequations}
\indent In problem (\ref{MPC_multi}), $k$ and $N$ refer to the time step and the MPC horizon length, respectively.
Let $\mathcal{M}:=\{1,2,...,M\}$ denote the set of all indices of a total of $M$ controlled vehicles.
The planning states and inputs trajectories over the MPC horizon $N$ are denoted by $\bm{p^i}=\{p^i_k\}_{k=0}^N,\forall i\in\mathcal{M}$ and $\bm{u_{p}^i}=\{u_{p,k}^i\}_{k=0}^{N-1},\forall i\in\mathcal{M}$, respectively.
The cost function in (\ref{MPC_multia}) penalizes the difference of planning states and reference planning position $p_r^i,\forall i\in\mathcal{M}$ and the planning inputs of all vehicles with weight matrices $Q$ and $R$, respectively. 
For the $i$th vehicle, $\mathcal{I}_p^i$ represents its initial planning state region, $\mathcal{G}_p^i$ represents its terminal planning state region, and $\mathcal{P}_{p,k}^i$ represents its time-varying bounded region for the planner along the prediction horizon.
The control inputs $u_{p,k}^i$ is subject to the input constraint $\mathcal{U}_p^i$. In the system dynamics constraint (\ref{MPC_multid}), the function $h_d(\cdot,\cdot)$ is the discrete model of the planning system.

When successfully solved, problem (\ref{MPC_multi}) generates collision-free trajectories for all the vehicles, simultaneously, guaranteed by the constraints (\ref{MPC_multie}) and (\ref{MPC_multif}), for obstacle and inter-vehicle collision avoidance, respectively.
Let $\mathcal{Q} := \{1,2,...,M_o\}$ denote the set of all indexes of the uncontrolled dynamical obstacles whose index is denoted by $q$. 
In the $n$-dimensional space, the set $\mathbb{O}^q\subseteq \mathbb{R}^n$ denotes the region occupied by the $q$th static obstacle, and $\mathbb{C}(p_k^i)\subseteq \mathbb{R}^n$ denotes the region occupied by the $i$th controlled vehicle at time step $k$. However, (\ref{MPC_multie}) and (\ref{MPC_multif}) are nonconvex nonsmooth constraints in general that make (\ref{MPC_multi}) hard to solve.

\subsection{Reformulation of the collision avoidance constraints}
One explicit representation of the collision avoidance constraints (\ref{MPC_multie}) and (\ref{MPC_multif}) is $\operatorname{dist}(\mathcal{A},\mathcal{B}) \geq 0$, requiring the minimum distance between any two convex sets $\mathcal{A}$ and $\mathcal{B}$ in a form of a convex cone to be greater than 0. 
The minimum distance problems $\operatorname{dist(\cdot,\cdot)}$ are optimization problems and cannot be used directly in the constraints of (\ref{MPC_multi}).
While this representation is still nonconvex nonsmooth in general, it can be reformulated as equivalent smooth nonconvex constraints via strong-duality theory inspired by \cite{b3}.

The reformulated collision avoidance constraints guarantee that the optimization problem (\ref{MPC_multi}) is a smooth nonlinear problem solvable by general-purpose nonlinear and nonconvex optimization solvers such as IPOPT \cite{b8}. 

We assume that the static obstacles $\mathbb{O}$ and controlled vehicles $\mathbb{C}(p)$ are presented by convex compact sets with the non-empty relative interior so that strong-duality holds \cite{b10}. The sets can be described by the following conic representations:
\begin{equation}
    \mathbb{O} = \{w\in\mathbb{R}^n: Aw\preceq_{\mathcal{K}_o}b\},
    \label{set_p}
\end{equation}
\begin{equation}
    \begin{aligned}
        \mathbb{C}(p) &=d(p)+R(p)\mathbb{E},\: \mathbb{E}=\{y\in\mathbb{R}^n:Gy\preceq_{\mathcal{K}_c}h\},
    \end{aligned}
    \label{set_e}
\end{equation}
where the matrices $(A,b)\in\mathbb{R}^{l\times n}\times\mathbb{R}^{l}$ associated with cone $\mathcal{K}_o\in\mathbb{R}^l$ define the shape of the obstacle, and matrices $(G,h)\in\mathbb{R}^{m\times n}\times\mathbb{R}^{m}$ associated with cone $\mathcal{K}_c\in\mathbb{R}^m$ define the shape of the vehicle. Both cones are assumed to be closed convex pointed cones with non-empty interior. 
Moreover, translation and rotation of the controlled vehicle, whose shape is modelled by the convex set $\mathbb{E}$, are modelled by the state-dependent translation vector $d:\mathbb{R}^{n_p}\to\mathbb{R}^n$ and the rotation matrix $R:\mathbb{R}^{n_p}\to\mathbb{R}^{n\times n}$, respectively.
% We refer to \cite{b3} for the discussion about the generalization of the dual reformulation. 

In this paper, we provide two concrete reformulations of collision avoidance constraints (\ref{MPC_multie}) and (\ref{MPC_multif}) by choosing $\mathcal{K}_o=\mathbb{R}_+^l$ to be polyhedral to describe an obstacle, and $\mathcal{K}_c$ to be a second-order cone to represent controlled vehicles in an ellipsoidal shape, which is a common shape for the AUVs.
Therein, $\preceq_{\mathcal{K}_o}$ is equivalent to the element-wise inequality $\leq$, and $Gy\preceq_{\mathcal{K}_c}h$ is represented by $\|P^{(-1/2)}y\|_2\leq1$, where the symmetric and positive-definite matrix $P\in\mathbb{R}^{n\times n}$ defines the scale of the ellipsoid.

\subsubsection{Obstacle collision avoidance}
\indent The Eulerian distance between the $q$th $\mathbb{O}^q$ and $i$th $\mathbb{C}(p^i)$ in the primal space is 
\begin{equation}
    \begin{aligned}
        &\mathrm{dist}(\mathbb{C}(p^i),\mathbb{O}^q)= \min_{w,y\in\mathbb{R}^n}\|(d(p^i)+R(p^i)y)-w\|_2, \\
        &\mathrm{s.t.} \: \|P^{(-1/2)}y\|_2\leq1, A^qw\leq b^q,
    \end{aligned}
    \label{primal_ep}
\end{equation}
whose dual problem is 
\begin{equation}
    \begin{aligned} 
    &\mathrm{dist}(\mathbb{C}(p^i),\mathbb{O}^q):=\max_{\lambda,\mu,z} -\lambda-\mu^\top b^q+z^\top d(p^i)\\
    &\mathrm{s.t.}\: \|(R(p^i)P^{(1/2)})^\top z\|_2\leq\lambda, (A^q)^\top\lambda-z=0, \|z\|_2\leq1
    \end{aligned}
    \label{dual_ep}
\end{equation}
where $\lambda\in\mathbb{R}$, $\mu\in\mathbb{R}^l$, and $z\in\mathbb{R}^n$ are dual variables of this problem. Therefore, the reformulation of constraint \eqref{MPC_multie}, which is $\mathrm{dist}(\mathbb{C}(p^i_k),\mathbb{O}^q)\geq0$, is as follows:
\begin{subequations}
\begin{align}
&-\lambda_k-\mu^\top_kb^q+z^\top_kd(p^i_k)\geq 0\\
&\|(R(p^i_k)P^{(1/2)})^Tz_k\|_2\leq\lambda,\\
&(A^q)^\top\lambda_k-z_k=0,\\
&\|z_k\|_2\leq1
\end{align}
\label{Refo_static}
\end{subequations}
\subsubsection{Inter-vehicle collision avoidance}
\indent The Eulerian distance between the controlled vehicles $\mathbb{C}(p^i)$ and $\mathbb{C}(p^j)$ in the primal space is 
\begin{equation}
    \begin{aligned}
        &\mathrm{dist}(\mathbb{C}(p^i),\mathbb{C}(p^j))\\
        =&\min_{y^i,y^j\in\mathbb{R}^n}\|(d(p^i)+R(p^i)y^i)-(d(p^j)+R(p^j)y^j))\|_2, \\
        &\mathrm{s.t.}\|P^{(-1/2)}y^i\|_2\leq1,\|P^{(-1/2)}y^j\|_2\leq1,
    \end{aligned}
    \label{primal_ee}
\end{equation}
whose dual problem is 
\begin{equation}
    \begin{aligned} 
    &\mathrm{dist}(\mathbb{C}(p^i),\mathbb{C}(p^j):=\max_{\lambda^i,\lambda^j,z}-\lambda^i-\lambda^j+z^\top(d(p^i) - d(p^j)), \\
    &\text{s.t.}\:\|(R(p^i)P^{(1/2)})^\top z\|_2\leq\lambda^i,\, \|(R(p^j)P^{(1/2)})^\top z\|_2\leq\lambda^j\,\\
    &\:\:\quad\|z\|_2\leq1,
    \end{aligned}
    \label{dual_ee}
\end{equation}
where $\lambda_i,\lambda_j\in\mathbb{R}$, and $z\in\mathbb{R}^n$ are dual variables of this problem. Therefore, the reformulation of constraint \eqref{MPC_multif}, which is $\mathrm{dist}(\mathbb{C}(p^i_k),\mathbb{C}(p^j_k))\geq0$, is as follows.
\begin{subequations}
\begin{align}
&-\lambda^i_k-\lambda^j_k+z_k^\top(d(p_k^i) - d(p_k^j))\geq 0, \\
&\|(R(p^i_k)P^{(1/2)})^\top z_k\|_2\leq\lambda^i_k, \\
&\|(R(p^j_k)P^{(1/2)})^\top z_k\|_2\leq\lambda^j_k, \\
&\|z_k\|_2\leq1.
\end{align}
\label{Refo_dynamic}
\end{subequations}

% The derivation of (\ref{dual_ep}) from (\ref{primal_ep}) and the derivation of (\ref{dual_ee}) from (\ref{primal_ee}) are provided in the Appendix.
\begin{remark}
The static polyhedral obstacle $\mathbb{O}$ and the initial vehicle shape $\mathbb{E}$ can also be formulated as \textbf{dynamic} ones which vary with time by representing $(A(t)$, $b(t))$ and $(G(t)$, $h(t))$ as functions of $t$. 
\end{remark}

\subsection{Quadrotor Example}
\begin{figure}[tbp]
\centerline{\includegraphics[width=8.5cm]{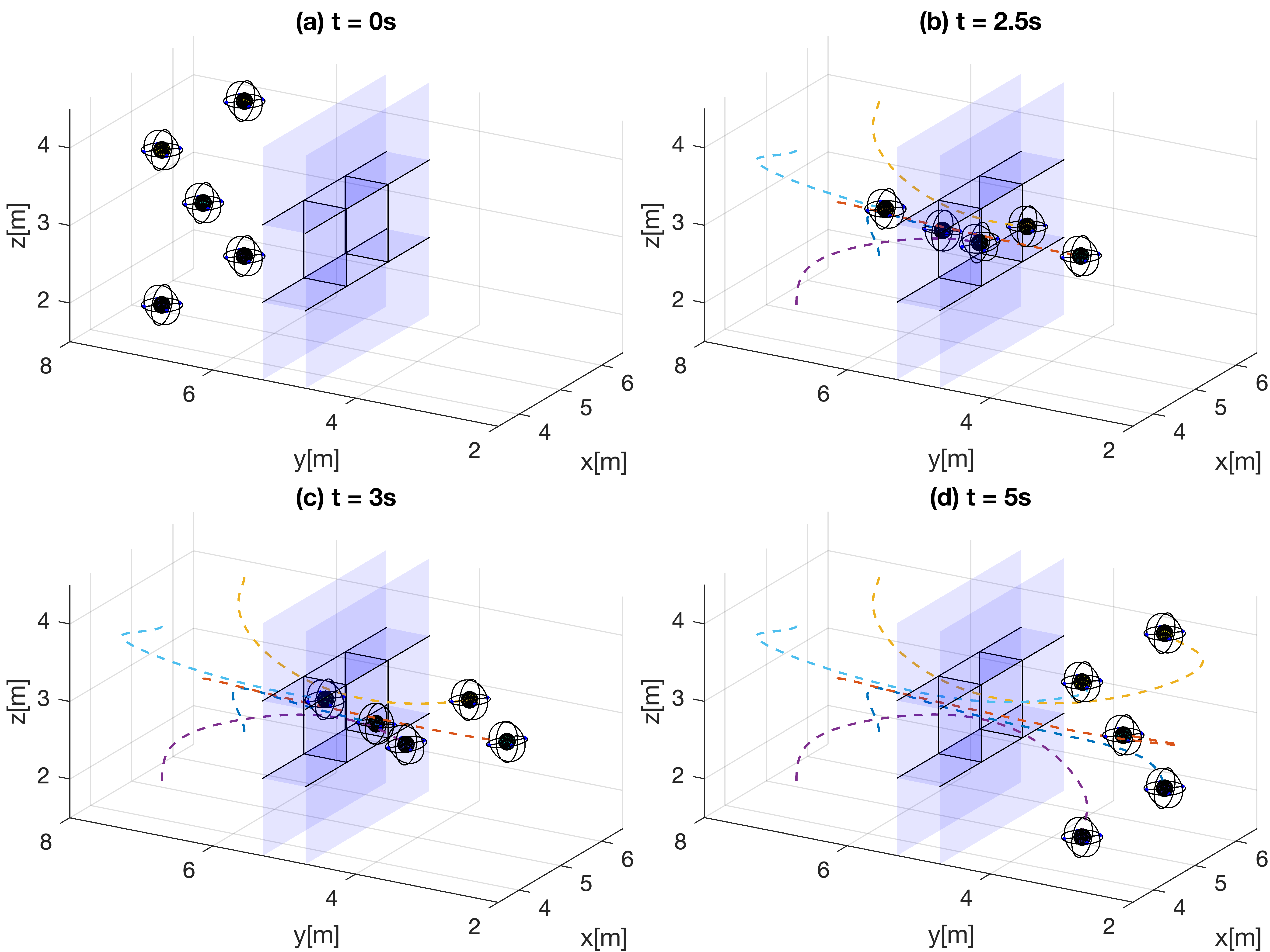}}
\caption{Trajectory planning for five 12D quadrotors}
\label{fig_quad}
\end{figure}

In this section, we demonstrate the proposed MPC trajectory planning for multiple vehicles in (\ref{MPC_multi}) on a quadrotor example. To allow the quadrotors to reach their respective goal positions while achieving static obstacle collision avoidance by passing through a small hole on the wall, and collision avoidance by keeping a safe distance between quadrotors, the trajectories of five quadrotors with the 12 DOF dynamic model in \cite{b5} are generated by solving the problem (\ref{MPC_multi}). 
Fig. \ref{fig_quad} shows trajectories and positions of the quadrotors at sampled time steps. The full simulation could be found at \url{https://youtu.be/K6sBlWbZYCM}.

States and inputs of multiple higher-dimensional systems and the dual variables introduced by the reformulated collision avoidance constraints make the problem (\ref{MPC_multi}) difficult to compute. The average computation time of the nonlinear solver IPOPT is 40 seconds for the five-second-long planning scenario shown in Fig. \ref{fig_quad}. The PC used has an i5-9300 CPU and 8GB RAM. Furthermore, model uncertainties and external disturbances are not considered, and safety cannot be guaranteed. To generate safety-guaranteed trajectories for multiple vehicles in real-time, we combine the MPC planning problem \eqref{MPC_multi} with the HJ differential game formulation.

\section{Safety-guaranteed trajectory planning}
We propose a novel safety-guaranteed online trajectory planning method for multi-AUV systems that experience wave disturbances, which also works for a single AUV.
The key idea is to, during planning, augment the AUV by the TEB derived in Section \ref{S_II_C}, considered as a safety bubble. 
Different from the earlier works \cite{b2} and \cite{b1}, in which the TEB is used to augment the environment and the uncontrolled obstacles for safety guarantees, our approach demonstrates obvious advantages in handling dynamical obstacles and reducing conservativeness in presence of many static obstacles, especially when the TVSD wave disturbance model is considered. 
Moreover, the shapes of the AUVs, together with the TEB, are considered in the planning problem to generate safety-guaranteed trajectories with respect to the augmented AUVs. 

Our approach is presented in a generalized fashion and can be divided into two phases. 
\begin{enumerate}
    \item The offline initialization phase first calculates the TEB by following the same procedures introduced in \cite{b1}, then augments the AUVs using the TEBs as safety bubbles with the suitable preprocessing based on the wave disturbance model chosen. Lastly, the constraints in (\ref{MPC_multi}) are initialized. 
    \item The online trajectory planning phase solves the MPC problem in (\ref{MPC_multi}) for safety-guaranteed trajectories for the augmented multi-AUV system. While solving one planning problem is enough for the system with the TVSI wave disturbance model, a re-planning strategy is required for the system with the TVSD wave disturbance model to iteratively reduce conservativeness in the trajectory.
\end{enumerate}
The planning problem (\ref{MPC_multi}) generates trajectories with respect to the augmented planning system. It is guaranteed that the tracking system can remain within a safe neighborhood of the trajectory while ensuring no collision happens.

\subsection{Trajectory Planning with TVSD Disturbance}
% We solve equation (\ref{Calculate_TEB}) through the helperOC toolbox in MATLAB for the TEB (\ref{TEB_SI}), and through the Optimized DP in Python for the TEB (\ref{TEB_SD}) to get discrete data sets. Therefore, we define the discrete representations of the TEB (\ref{TEB_SI}) by $\mathcal{B}_{e,1}(k)$ and the TEB (\ref{TEB_SD}) by $\mathcal{B}_{e,2}(x_d,z_d,k)$

For the offline initialization phase, we first calculate the TEB defined in (\ref{TEB_SD}) by using (\ref{Calculate_TEB}) and following the same procedures in \cite{b1}. 
Note that the TEB defined in (\ref{TEB_SI}) with the TVSI wave disturbance model $\mathcal{B}_{e,1}(t)$ only depends on time $t$ whilst the TEB defined in (\ref{TEB_SD}) with the TVSD wave disturbance model $\mathcal{B}_{e,2}(x,z,t)$ depends not only on time $t$, but also on the position of the tracking system. The difference between the two kinds of TEBs motivates us to propose a novel generalized trajectory planning algorithm. 

Due to limited computation resources online, we want to approximate $\mathcal{B}_{e,2}(x,z,t)$ for each $(x,z,t)$ pair as a ball set, whose radius is stored in a 3D look-up table for real-time computation.
Specifically, the approximation is done by using uniform grids to partition the $x$-$z$ plane, where the center of each grid is denoted by $(x_d,z_d)$. 
Also, the planning time is discretized as $t_k=kT_s=kT/N$, where $T_s$ is the discretization interval and $T$ is the total planning time for the planner. 
The 3D look-up table stores the radius $\mathcal{R}(x_d,z_d,t_k)$ of the outer approximation of the TEB $\mathcal{B}_{e,2}(x,z,t)$ for each $(x_d,z_d,t_k)$ pair, where the radius is calculated offline as follows:
\begin{equation}\begin{aligned}
    &\mathcal{R}(x_d,z_d,t_k)\\
    &=\max_{e\in\mathcal{B}_{e,2}(x_d,z_d,t_k)}\|e\|_2,\forall t_k\in[0,T],[x_d,z_d]^\top\in\mathcal{P},
    \label{approx_B2}
\end{aligned}\end{equation}
where $\mathcal{P}$ denotes the feasible region of the tracking system. At the cost of larger storage requirement and higher online look-up computation cost, reducing the width of the grids and $T_s$ results in higher resolution of the 3D look-up table, leading to a better approximation of the true TEB. We next provide the approximated TVSD TEB as follows.
\begin{equation}\begin{aligned}
    &\mathcal{B}_{max,2}(t_k), \forall t_k\in[0,T]\\
    &=\{\beta\in\mathbb{R}^2: \|\beta\|_2\leq\max_{[x_d, z_d]^\top\in \mathcal{T}_k}\mathcal{R}(x_d,z_d,t_k)\},
    \label{max_B2}
\end{aligned}\end{equation}
where $\mathcal{T}_k$ is the region occupied by the augmented vehicles at each time step. For the first online planning iteration, the approximated TEB in \eqref{max_B2} used to augment AUVs is initialized by the maximum radius of each time step inside the look-up table, i.e. the most conservative case.
With $\mathcal{B}_{max,2}(t_k)$ only depending on $t_k$, we obtain $\mathbb{C}^a(p_k)$ to represent the augmented AUVs at each time step for problem \eqref{MPC_multi}:
\begin{equation}
    \mathbb{C}^a(p_k) =  \mathbb{C}(p_k)\oplus\mathcal{B}_{max,2}(t_k).
    \label{TVSD_augcar}
\end{equation}
Then the initial and goal feasible position sets in the planning are (\ref{Plan_I}) and (\ref{Plan_G}), respectively: 
\begin{equation}
    %\mathcal{I}^p =\{p_0:\mathbb{C}^a(p_0,0)\subseteq\mathcal{P}\},
     \mathcal{I}_p = \mathcal{P} \ominus \mathbb{C}^a(p_0,0),
    \label{Plan_I}
\end{equation}
\begin{equation}
    %\mathcal{G}^p =\{p_N:\mathbb{C}^a(p_N,T)\subseteq\mathcal{G}\}.
    \mathcal{G}_p =\mathcal{G} \ominus \mathbb{C}^a(p_N,T),
    \label{Plan_G}
\end{equation}
where $\mathcal{G}$ denotes the goal region of the tracking system. Similarly, the bounded feasible position set in the planning is:
\begin{equation}
    %\mathcal{P}^p(t_k) =\{p:\mathbb{C}^a(p_k,t_k)\subseteq\mathcal{P}\}.
    \mathcal{P}_{p,k} = \mathcal{P} \ominus \mathbb{C}^a(p_k,t_k).
    \label{Plan_P}
\end{equation}
\begin{remark}
Since the rotation of the AUVs is not considered, $\mathbb{C}^a(p_k,t_k)$ is directly used on the right-hand-side of $\ominus$. If the rotation is considered, a ball set as an outer approximation of $\mathbb{C}^a(p_k,t_k)$ should be used instead. 
\end{remark}
\begin{algorithm}[b]
\caption{Online Updating of the MPC problem in \eqref{MPC_multi}}\label{algorithm_2}
\begin{algorithmic}[1]
\STATE \textbf{Input:} $\mathcal{B}_{approx,2}(x_d,z_d,t_k)$, $\mathbb{C}^a(p^i_k)$, $\forall k \in [0,N]$, $i \in [1,M]$
\FOR{$k \in [0,N]$}
\FOR{$i \in [1,M]$}
% \STATE Find $\mathcal{R}_{k}^i$  occupied by
% $\mathbb{C}^a(p^i_k)$
\STATE Update $\mathcal{B}_{max,2}^i(t_k)$ using (\ref{max_B2}) with $\mathbb{C}^a(p^i_k)$
\STATE Update $\mathbb{C}^a(p^i_k)$ using (\ref{TVSD_augcar}) with $\mathcal{B}_{max,2}^i(t_k)$
\STATE Update $\mathcal{I}^i_p$ using (\ref{Plan_I}) with $\mathbb{C}^a(p^i_k)$
\STATE Update $\mathcal{G}^i_p$ using (\ref{Plan_G}) with $\mathbb{C}^a(p^i_k)$
\STATE Update $\mathcal{P}^i_{p,k}$ using (\ref{Plan_P}) with $\mathbb{C}^a(p^i_k)$
\ENDFOR
\ENDFOR
\STATE \textbf{Output:} $\mathbb{C}^a(p^i_k)$, $\mathcal{I}_p^i$, $\mathcal{G}_p^i$, $\mathcal{P}^i_{p,k}$
\end{algorithmic}
\end{algorithm}

For the online trajectory planning phase, the main step is to solve the MPC problem in (\ref{MPC_multi}) with constraints \eqref{MPC_multie} and \eqref{MPC_multif} reformulated as \eqref{Refo_static} and \eqref{Refo_dynamic}, respectively. However, to reduce conservativeness of the trajectories planned, we need to update $\mathcal{B}_{max,2}(t_k)$ by finding the region occupied by the augmented vehicles from the previous planned trajectories. We set $\mathcal{T}_k = \mathbb{C}^a(p_k)$ after the first planning iteration. Then, $\mathcal{B}_{max,2}(t_k)$ is updated by (\ref{max_B2}). The augmented controlled vehicles (\ref{TVSD_augcar}) and feasible sets (\ref{Plan_I}-\ref{Plan_P}) are updated in real-time subsequently.

Let $l\in[1,\dots,l_{max}]$ denote the iteration index of the online planning with a maximum iteration number $l_{max}$. Also, let $\xi$ be the solution of the MPC problem in \eqref{MPC_multi}. The planning time depends on the re-planning time and the complexity of the problem. Using the solution $\xi_{l-1}$ obtained from the previous MPC problem as the warm-start $\xi_{warm}$ of the current iteration, the computation time can be reduced. We summarize the online update of constraints in Algorithm \ref{algorithm_2} and the procedure of trajectory planning with TVSD wave disturbance model in Algorithm \ref{algorithm_1}.

\begin{algorithm}[h]
\caption{Safety-Guaranteed Trajectory Planning for Multiple AUVs under State-Dependent Wave Disturbance}\label{algorithm_1}
\begin{algorithmic}[1]
\STATE \textbf{Input:} $\mathbb{O}_q$, $q\in[0,M_o]$, $T$, $N$, $f(s^i,u_s^i,d,t)$, $h(p^i,u_p^i)$, $i\in[1,M]$
\STATE \textbf{Offline initialization:} 
\STATE Derive $g_2(r_2,u_s,u_p,d,t)$, $t\in[0,T]$ using (\ref{relative_SD})
\STATE Calculate $\mathcal{B}_{e,2}(x,z,t)$ using (\ref{min_V}) and (\ref{TEB_SD}) based on $g_2(r_2,u_s,u_p,d,t)$
\STATE Build the 3D look-up table to store the radius of the approximation of the TEB, where the radius $\mathcal{R}(x_d,z_d,t_k)$ is calculated by (\ref{approx_B2})
\STATE Find $\mathcal{B}_{max,2}^i(t_k)$, $\forall t_k\in[0,T]$, $i\in[1,M]$ using (\ref{max_B2})
\STATE Calculate $\mathbb{C}^a(p^i_k)$, $\forall k\in[0,N]$, $i\in[1,M]$ using (\ref{TVSD_augcar})
% \ENDFOR
\STATE Calculate $\mathcal{I}_p^i$, $\forall i\in[1,M]$ using (\ref{Plan_I})
\STATE Calculate $\mathcal{G}_p^i$, $\forall i\in[1,M]$ using (\ref{Plan_G})
\STATE Calculate $\mathcal{P}^i_{p,k}$, $\forall k\in[0,N]$, $i\in[1,M]$ using (\ref{Plan_P})
\STATE Find the discrete model $h_d(p,u)$
%using Euler discretization with $T_s$
\STATE Initialze $\xi_{0}$ with a column vector of zeros
\STATE \textbf{Online trajectory planning}
\FOR {$l$ $\in [1,\dots, l_{max}]$} 
\STATE Set warm-start $\xi_{warm}$ to $\xi_{l-1}$
\STATE Dervie solution $\xi_l$ by solving MPC problem in (\ref{MPC_multi}) with $h_d(p,u)$, $\mathbb{C}^a(p^i_k)$, $\xi_{warm}$, $\mathcal{I}_p^i$, $\mathcal{G}_p^i$, $\mathcal{P}^i_k$, $k\in[0,N]$ and $i\in[1,M]$, and with \eqref{MPC_multie} and \eqref{MPC_multif} reformulated as \eqref{Refo_static} and \eqref{Refo_dynamic}, respectively.
\STATE Update constraints of MPC problem by Algorithm \ref{algorithm_2}
\ENDFOR
\STATE \textbf{Output:} trajectory $\bm{p}^i$, $\forall i\in[1,M]$
\end{algorithmic}
\end{algorithm}
\subsection{Trajectory Planning with TVSI Disturbance}\label{S_IV_A}

The algorithm to plan trajectories with TVSI wave disturbance model can be derived by simplifying Algorithm \ref{algorithm_1}. Specifically, in the offline phase, we use $g_1(r_1,u_s,u_p,d,t)$, $t\in[0,T]$ in step 3 and calculate $\mathcal{B}_{e,1}(t)$ through (\ref{min_V}) and (\ref{TEB_SI}) based on $g_1$ in step 4. Step 5 and 6 are dropped because the TEB with TVSI disturbance $\mathcal{B}_{e,1}(t)$ is only time-dependent. The controlled vehicles augmented by the TEB under the TVSI disturbance can be defined as follows:
\begin{equation}
    \mathbb{C}^a(p_k) =  \mathbb{C}(p_k)\oplus\mathcal{B}_{e,1}(t_k),
    \label{TVSI_augcar}
\end{equation}

Because the TEB does not change with the states, the MPC problem \eqref{MPC_multi} is only solved for one time. By setting $l_{max} = 1$ in step 14, no constraint update is required in step 17 and Algorithm \ref{algorithm_2} becomes easy to calculate.

% For both cases, the most time-consuming part, calculation of the TEB, is achieved offline through the use of a 3D look-up table. 
\begin{remark}\label{remark_4}
Algorithm \ref{algorithm_1} can be modified for single vehicle trajectory planning with various wave disturbance models by setting $M = 1$ and dropping constraint \eqref{MPC_multif} in the MPC problem \eqref{MPC_multi}.
\end{remark}

\section{Simulation}
% In this section, the proposed algorithms are first applied to single AUV trajectory planning problems with the three different wave disturbance models shown in \cite{b1} to successfully generated feasible trajectories.
% Then we provide one more single AUV simulation with TVSD wave disturbance model and a extreme obstacle setting to show that our approach is less conservative than that in \cite{b1}.
% Finally, a multi-AUV simulation under TVSD wave disturbance is used to demonstrate the ability of our approach to find less conservative trajectories with reduced computation time.

In this section, the proposed algorithms are first applied to single AUV trajectory planning problems with the wave disturbance models introduced in Section \ref{S_II_B} to compare our approach to that proposed in \cite{b1}.
Then, a multi-AUV simulation under TVSD wave disturbance is presented to demonstrate the ability of our approach to find less conservative trajectories, through iterative replanning, that reaches the goal region faster.

\subsection{Single AUV planning under various wave disturbance models}\label{S_V_A}
We compare the effect of the proposed algorithms applied to the three cases introduced in \cite{b1}. Parameters of the AUV model (\ref{AUV_model}) are $m=116$kg, $\Bar{m}=116.2$kg, $X_{\Dot{u}_r}=-167.7$, $Z_{\Dot{w}_r}=-383$, $X_{u_r}=26.9$, $Z_{w_r}=0$, $X_{|u_r|u_r}=241.3$, and $Z_{|w_r|w_r}=265.6$, same as in \cite{b9}. The bounded system disturbances are given as $d_x,d_z,d_{u_r},d_{w_r}\in[-0.001,0.001]$. The tracking inputs and planning inputs are constrained by $T_A,T_B\in[-70,70]$ and $u_{p,x},u_{p,z}\in[-0.02,0.02]$, respectively.

The parameters of the TVSD disturbance model in (\ref{TVSD_V}) and (\ref{TVSD_A}) are $A=0.05\mathrm{m}$, $\omega=\pi\mathrm{rad/s}$, and $k=\omega^2/g=1.0061\mathrm{rad/m}$. The parameters of the TVSI disturbance model in (\ref{TVSI_V}) and (\ref{TVSI_A}) are $\Tilde{A}_v = 0.0167$, $\Tilde{A}_a = 0.0523$, and $\Tilde{\phi}_v=\Tilde{\phi}_a=0$ with additional bounded disturbance $d_{f,v,x},d_{f,v,x}\in[-0.007,0.007]$ and $d_{f,a,x},d_{f,a,x}\in[-0.02,0.02]$. 

We further introduce the time-invariant state-independent (TISI) wave disturbance model, which are the smallest time-invariant bounds of the TVSD disturbance model in \eqref{TVSD_V} and \eqref{TVSD_A} for $[x,z]^\top\in\mathcal{P}$ and $t\in[0,T]$. The bounds are selected with  $V_{f,x},V_{f,z}\in[-0.021,0.021]$, and $A_{f,x},A_{f,z}\in[-0.0660,0.0660]$.
Notice that the relative system (\ref{relative_SI}) can also be used with the TISI wave disturbance model. We refer to \cite{b1} for more details about the parameters of the three wave disturbance models as well as the TEB generation process using the above setting.
\begin{figure}[tbp]
\centerline{\includegraphics[width=8.5cm]{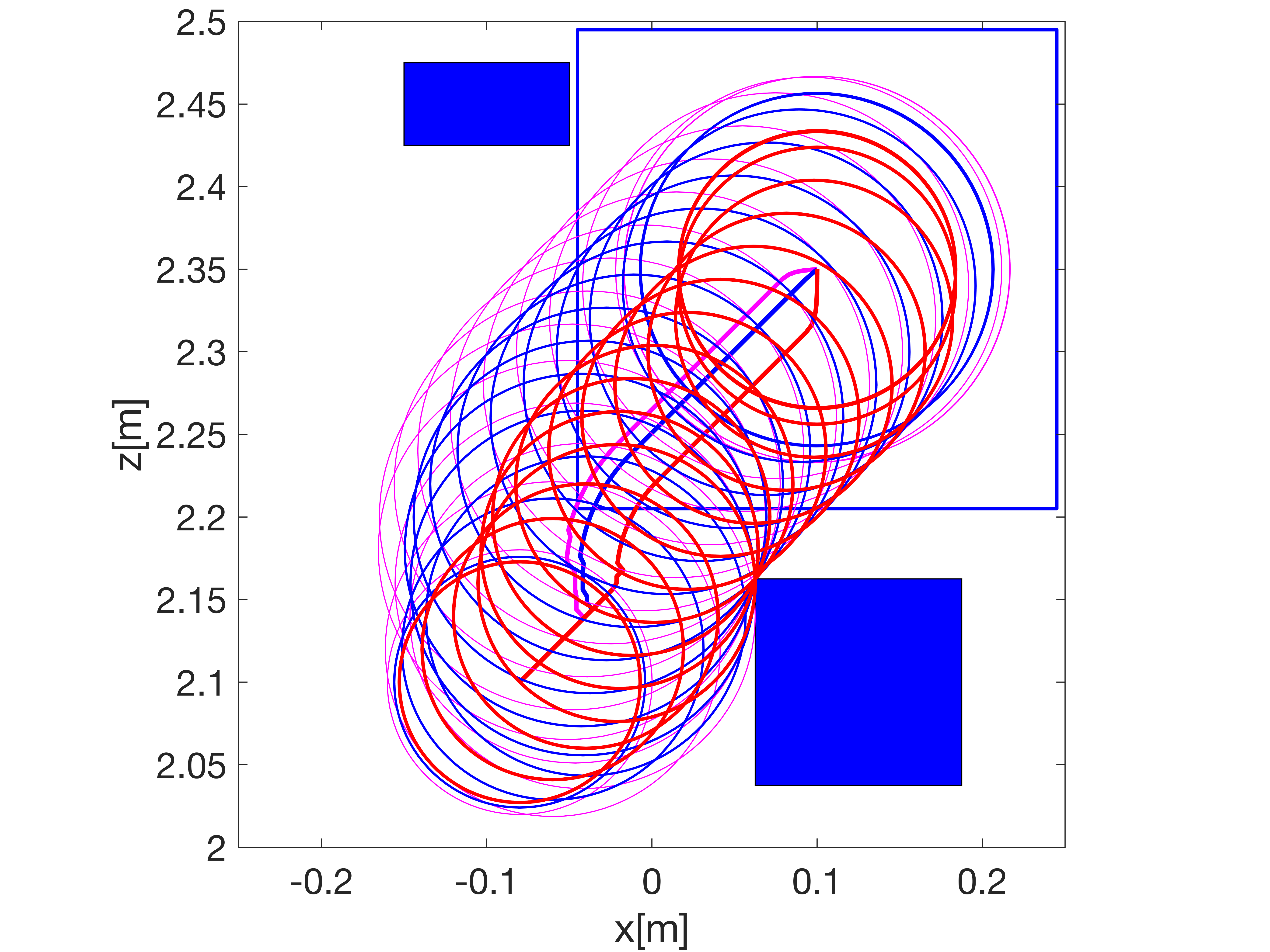}}
\caption{Planned trajectories (lines) and occupied region (circles) by the augmented vehicles of the three cases. Therein, the line and circle with color (\color{red}--\color{black}) indicate the converged case with TVSD wave disturbance model, the line and circle with color (\color{blue}--\color{black}) indicate the case with TVSI wave disturbance model, and the line and circle with color (\color{magenta}--\color{black}) indicate the case with TISI wave disturbance model. The 
solid blue rectangles represent obstacles, and the empty blue box represents the goal region.}
\label{fig_S1}
\end{figure}

For all three cases, the initial planning position is $[x_{p,0},z_{p,0}]^\top=[-0.08,2.1]^\top$. The feasible region is $\mathcal{P}=\{[x,z]^\top:x\in[-0.25,0.25],z\in[2,2.5]\}$, and the goal region is $\mathcal{G}=\{[x,z]^\top:x\in[-0.045, 0.245],z\in[2.205,2.495]\}$ with the centre position of the goal region $p_{r} = [0.1,2.35]^\top$ as the reference position in problem (\ref{MPC_multi}). The time interval for planning is $t\in[0,20]$s with a sampling period $T_s=0.2$s.

Modeling the vehicle as a point-mass, we apply Algorithm \ref{algorithm_1} with constraint \eqref{MPC_multif} dropped and constraint \eqref{MPC_multie} reformulated to solve the trajectory planning problem (see Remark \ref{remark_4}).
We illustrate the trajectories (represented by lines) of the AUVs and the regions occupied by the augmented AUVs (represented by circles) in Fig. \ref{fig_S1}. It can be observed that the case with the TVSD wave disturbance model leads to the least conservative trajectory with the smallest safety bubble and takes 9.8 seconds to reach the goal region. The case with the TVSI wave disturbance model takes 11 seconds to reach the goal region, and the case with the TISI wave disturbance model, being the most conservative, takes 15s to reach the goal region. 
% Notice that the resolution of the TEB look-up table is limited by computation resources offline. Thus, the TEB with the TVSD wave disturbance model only varies slightly between different discrete positions and time steps. 
After the second planning iteration (Step 14-18 in Algorithm 2\ref{algorithm_1}), the optimal trajectory was reached, with no further reduction in TEB size. 
% The total computation time for two planning iterations is 3 seconds, which is suitable for real-time planning. 
\begin{figure}[tbp]
\centerline{\includegraphics[width=8.5cm]{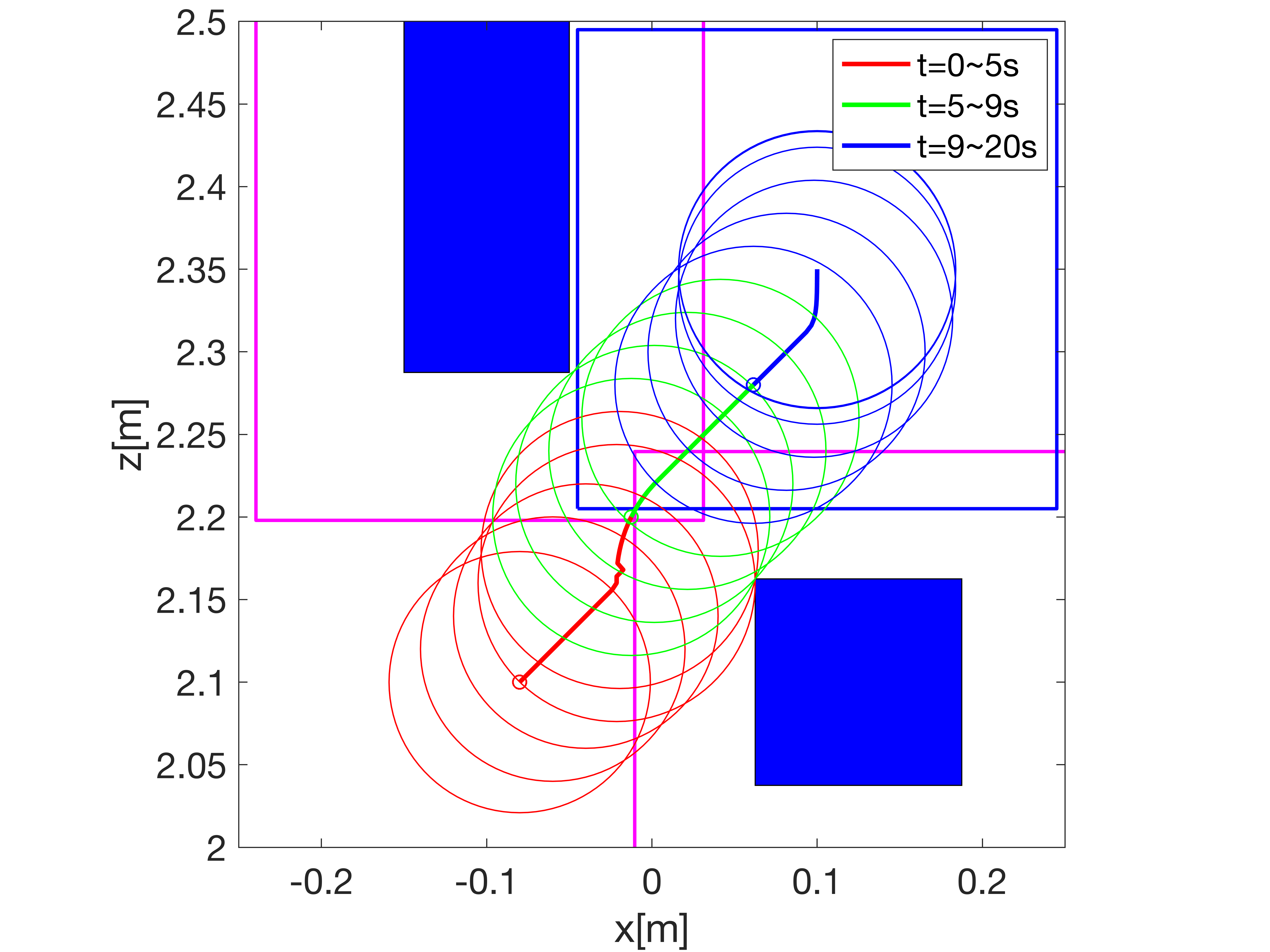}}
\caption{The augmented obstacles (empty rectangles with color (\color{magenta}--\color{black})) derived in \cite{b1} overlap at $t=5s$, before the AUV pass them. However, it is still feasible for the AUV solved by the algorithm in this paper, where the lines and circles with different color represent the feasible planned trajectory and the occupied region at different time period. Obstacles and goal region keep the same definition as Fig. \ref{fig_S1}}
\label{fig_S2}
\end{figure}

Our approach shows clear advantages in the case presented in Fig. \ref{fig_S2}, with the same environment setting but larger obstacles. Using the algorithm in \cite[Algorithm 2]{b1} with the TVSD wave disturbance model, the obstacles are augmented to a level where no feasible trajectory exists. Our approach, on the other hand, considers a larger feasible region and hence provides a feasible trajectory.

\subsection{Multiple AUV planning under TVSD wave disturbance}
Here, we consider the trajectory planning for three ellipsoidal AUVs with the TVSD wave disturbance model. 
Let the AUVs have ellipsoid shapes described in \eqref{primal_ep} with $P = \mathrm{diag}(0.12,0.03)$.
We continue to use the planning system in \eqref{planning_system} and the tracking system in \eqref{AUV_model} in this simulation. Since the pitch angle is ignored, the rotation matrix is identity matrix $R(p)=I_2$ for the ellipsoidal sets that represent the AUVs.
The tracking inputs and planning inputs for all AUVs constrained by $T_A\in[-150,150]$, $T_B\in[-210,210]$ and $u_{p,x}\in[-0.05,0.05]$, $u_{p,z}\in[-0.03,0.03]$, which is different from the previous simulation to accommodate for the larger environment. 
% To achieve safety guarantees, the tracking input $T_B$ in $z$ direction is set to be larger while the planning input $u_{p,z}$ is made to be smaller compared with the other inputs in the corresponding systems because the wave disturbance (the same wave parameters as Section \ref{S_V_A}) is more sensitive in the $z$ direction.

For all AUVs, the feasible region is $\mathcal{P}=\{[x,z]^\top:x\in[-0.75,0.75],z\in[1.5,3]\}$, and the goal region is $\mathcal{G}=\{[x,z]^\top:x\in[0, 0.75],z\in[1.5,2.5]\}$ with the centre position of the goal region $p_{r}= [0.375,2]^\top$ as the reference position in problem (\ref{MPC_multi}). 
The initial positions for the three AUVs are $[x_{p,0}^1,z_{p,0}^1]^\top=[-0.35,1.95]^\top$, $[x_{p,0}^2,z_{p,0}^2]^\top=[-0.45,1.65]^\top$, and $[x_{p,0}^3,z_{p,0}^3]^\top=[0.35,2.8]^\top$, respectively.

\begin{figure}[bp]
\centerline{\includegraphics[width=8.5cm]{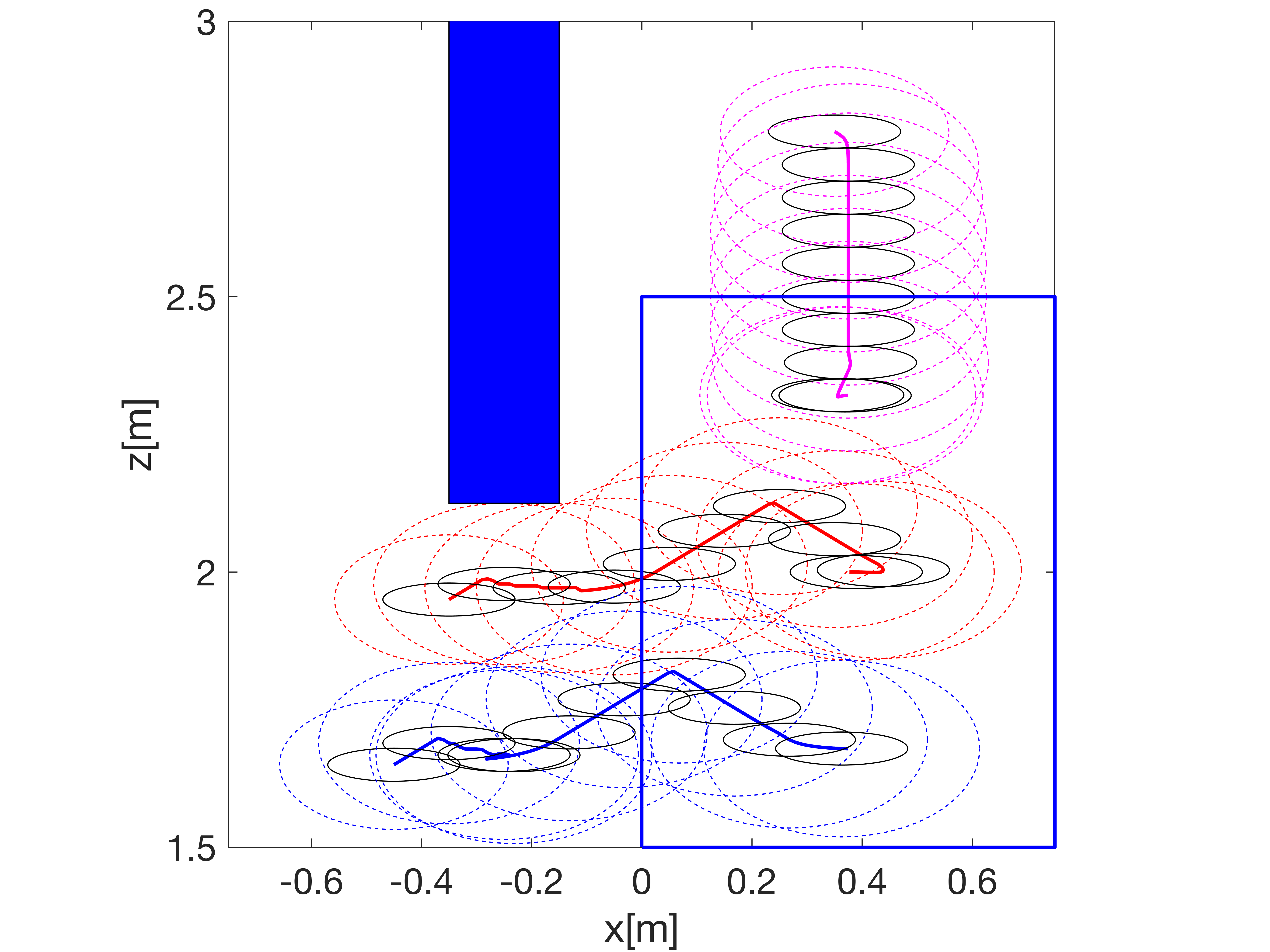}}
\caption{The third planning iteration: ellipsoids and the line with color (\color{red}--\color{black}) represent the augmented shape and the planned trajectory of the first AUV, ellipsoids and the line with color (\color{blue}--\color{black}) represents the augmented shape and the planned trajectory of the second AUV, and ellipsoids and the line with color (\color{magenta}--\color{black}) represents the augmented shape and the planned trajectory of the third AUV. Ellipsoids with color (\color{black}--\color{black}) represents the original shape of AUVs, The solid rectangle with color (\color{blue}--\color{black}) represents the obstacle and box with color (\color{blue}--\color{black}) represents the goal region.}
\label{fig_M1}
\end{figure}

\begin{figure}[hbt]
\centerline{\includegraphics[width=8.5cm]{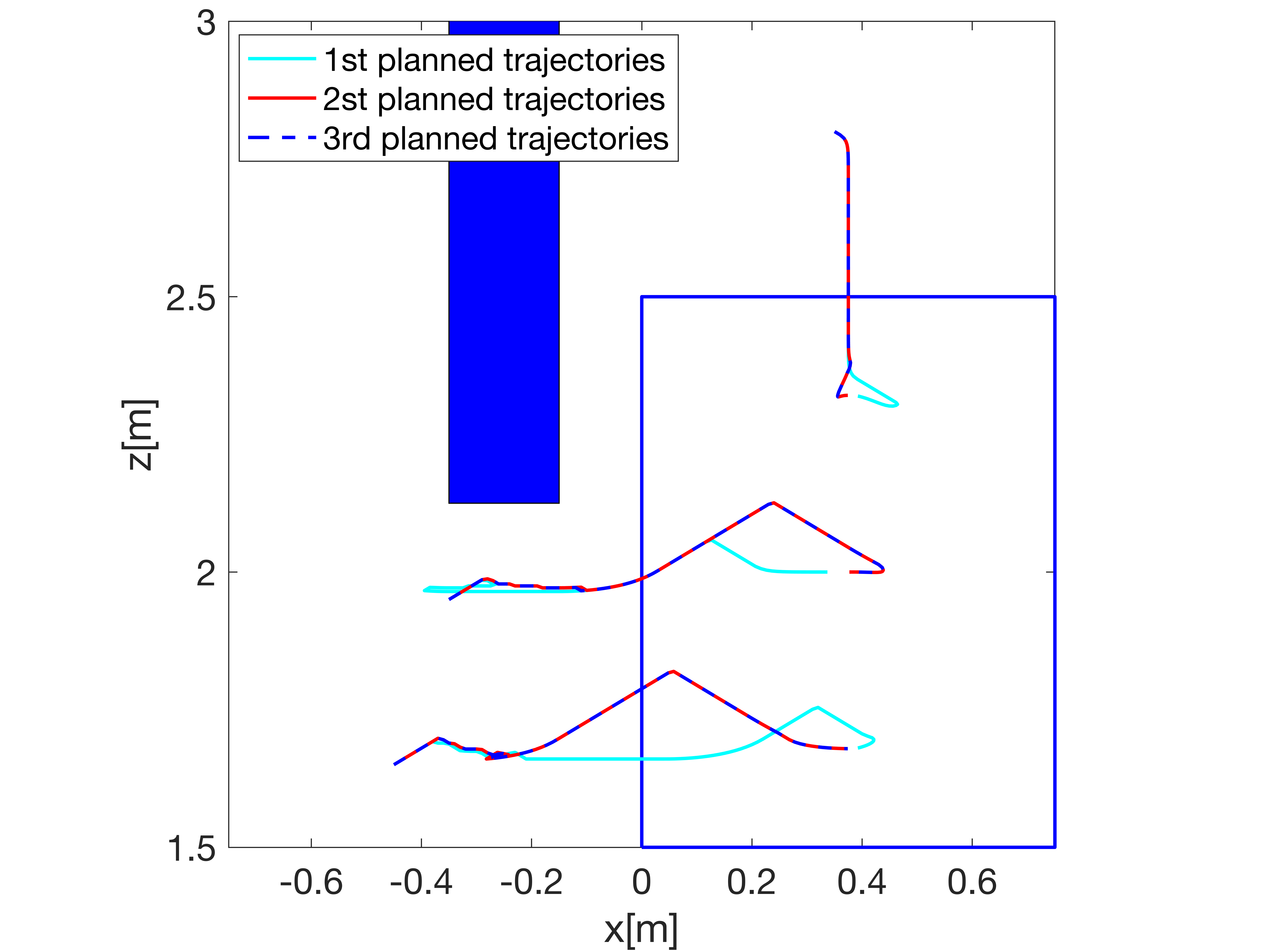}}
\caption{Trajectories of the first three planning iterations with the same environment defined in Fig.\ref{fig_M1}.}
\label{fig_M2}
\end{figure}

We apply Algorithm \ref{algorithm_1} with \eqref{MPC_multie} and \eqref{MPC_multif} in \eqref{MPC_multi} reformulated by \eqref{Refo_static} and \eqref{Refo_dynamic}, respectively, to solve the problem. Under this condition, the optimal trajectories are obtained at the third iteration in the for loop, i.e., step 14-18 in Algorithm \ref{algorithm_1} without further reduction in the augmented vehicle size. Fig.\ref{fig_M1} shows the trajectories of the three AUVs from the third planning iteration, which takes 15.8 seconds for all AUVs to reach the goal region. 
Fig. \ref{fig_M2} shows the planned trajectories at different iterations, i.e, $l=1,2,3$, and there is no obvious difference between the second and the third planned trajectories.
Compared with the trajectories from the second and the third iteration, 
the planned trajectories from the first iteration are more conservative, and hovering behavior in the first AUV (waiting for the other AUV to pass) at the start of the trajectory is observed. In the first iteration, it takes 18 seconds for all AUVs to reach the goal region, longer than that of the trajectories from the third iteration. 
Simulation of the multi-AUV trajectory planning could be found at \url{https://youtu.be/1fh1o3Gviik}. 

% With the previous solution of the MPC problem used to warm start the current planning iteration, the computation time can be shortened from 10.1 seconds for the first iteration, to 3.1 seconds for the second planning, then to 2.9 seconds for the third iteration. 

\end{document}